\documentclass[preprint,showpacs,preprintnumbers,amsmath,amssymb,superscriptaddress]{revtex4}

\usepackage[dvipdfm]{graphicx}
\usepackage{dcolumn}
\usepackage{bm}
\usepackage{amsmath, subfigure}

\begin{document}

\preprint{}

\title{Theory of Interaction of Memory Patterns in Layered Associative Networks}

\author{Kazuya Ishibashi}
 \email{kazuya@mns.k.u-tokyo.ac.jp}
\affiliation{Graduate School of Frontier Sciences, The University of Tokyo, Japan}
\affiliation{Japan Science and Technology Agency}
\author{Kosuke Hamaguchi}%
\affiliation{Brain Science Institute, RIKEN}
\author{Masato Okada}
\affiliation{Graduate School of Frontier Sciences, The University of Tokyo, Japan}
\affiliation{Japan Science and Technology Agency}
\affiliation{Brain Science Institute, RIKEN}

\date{\today}

\begin{abstract}
A synfire chain is a network that can generate repeated spike patterns with millisecond precision.  Although synfire chains with only one activity propagation mode have been intensively analyzed with several neuron models, those with several stable propagation modes have not been thoroughly investigated.  By using the leaky integrate-and-fire neuron model, we constructed a layered associative network embedded with memory patterns. We analyzed the network dynamics with the Fokker-Planck equation.  First, we addressed the stability of one memory pattern as a propagating spike volley.  We showed that memory patterns propagate as pulse packets.  Second, we investigated the activity when we activated two different memory patterns.  Simultaneous activation of two memory patterns with the same strength led the propagating pattern to a mixed state.  In contrast, when the activations had different strengths, the pulse packet converged to a two-peak state.  Finally, we studied the effect of the preceding pulse packet on the following pulse packet. The following pulse packet was modified from its original activated memory pattern, and it converged to a two-peak state, mixed state or non-spike state depending on the time interval.
\end{abstract}

\pacs{Valid PACS appear here}
\maketitle

\section{Introduction}
How our brains encode information is one of the most fascinating questions in neuroscience.  Some researchers expect that repeated spike patterns, which are observed repeatedly with millisecond precision, play a key role in the functioning of the cerebral cortex.  Such precise patterns are observed \textit{in vivo}~\cite{abeles93, abeles98} and \textit{in vitro}\cite{ikegaya}.  The synfire chain~\cite{corticonics} model is regarded as a model for generating repeated spike patterns.  A synfire chain is a functionally but not anatomically feed-forward network, and it transmits synchronous spikes called a pulse packet.  Because of the synchrony, it can reproduce spike patterns with millisecond precision.  Synfire chains have been intensively studied theoretically,~\cite{diesmann,cateau} and have been confirmed to exist \textit{in vivo} in an iteratively constructed network~\cite{reyes}.

Many of the studies on synfire chains use a homogeneous network structure, and the synfire chains have only one stable propagation mode, i.e., spatially uniform synchronized activity.  In contrast, synfire chains with several stable propagation modes have not been thoroughly investigated~\cite{hamaguchi,aviel}.  In particular, activity when network is driven toward several propagation modes simultaneously or successively has yet to be investigated.

In this paper, we construct a layered associative network composed of identical leaky integrate-and-fire neurons, and embed memory patterns into it.  Assuming that the number of neurons is large enough, we describe the membrane potential distribution with the Fokker-Planck equation and analyze the network dynamics.

Section \ref{sec2} explains the details of our layered associative network, and \S \ref{sec3} explains the Fokker-Planck method.  Section \ref{sec4} describes the results of our analysis.  In \S \ref{sec41}, we address the stability of single memory pattern propagation, and then investigate the interaction of two memory patterns by using simultaneous (\S \ref{sec421}) or successive (\S \ref{sec422}) activation.  Section \ref{sec5} is a summary and discussion.
\section{Layered Associative Network}\label{sec2}
\begin{figure}[tb]
\centering
\includegraphics[width=6cm]{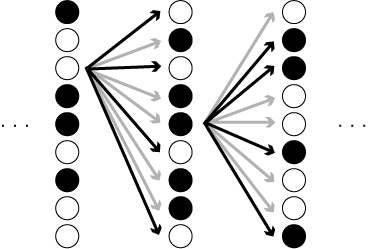}
\caption{Schematic diagram of our layered associative network.  Each circle means a neuron.  The color shows one memory pattern.  Black means $\xi^{l,\mu}_i=+1$, and white $-1$.  Arrows show synaptic connections.  Black lines means excitatory connections, and gray ones inhibitory.}
\label{fig_intro}
\end{figure}

In this paper, we consider a layered associative memory network, in which a synaptic connection on layer $l$ is given by
\begin{align}
J^l_{ij} = \frac{1}{N} \sum^p_{\mu=1} \xi^{l+1, \mu}_j \xi^{l, \mu}_i,
\end{align}
where the number of neurons in a layer is $i, j = 1, \ldots, N$ and the number of the memory patterns is $\mu=1,\ldots,p$.  $\xi^{l,\mu}_i$ represents the $\mu$th embedded memory pattern of the $i$th neuron on layer $l$ and takes on a value of either $+1$ or $-1$ according to the probability,
\begin{align}
\mathrm{Prob}[\xi^{l,\mu}_i = \pm 1] = \frac{1}{2}.
\end{align}
Figure~\ref{fig_intro} is a schematic diagram of this network.

We use the leaky integrate-and-fire (LIF) neuron model, and the dynamics of the membrane potential $v^l_i(t)$ can be described as a stochastic differential equation,
\begin{align}
\frac{{\rm d} v^l_i(t)}{{\rm d} t} = - \frac{v^l_i(t)-V_{\mathrm{rest}}}{\tau} + \frac{I^{l, \alpha}_i(t) + I_0}{C} + D' \eta(t),\label{eq_lif}
\end{align}
where $\tau$ is the membrane time constant, $V_{\mathrm{rest}}$ is the resting potential, $I_0$ is the mean of noisy input, $\eta(t)$ is white Gaussian noise satisfying $<\eta(t)>=0$ and $<\eta(t)\eta(t')>=\delta(t-t')$, and $D'$ is the amplitude of the noise.  Input current $I^{l,\alpha}_i(t)$ is obtained by convoluting the presynaptic firing $I^l_i(t)$ with the $\alpha$ function $\alpha(t) = \alpha^2 t \exp (-\alpha t)$.
\begin{align}
I^{l,\alpha}_i(t) &= \beta \int^{\infty}_0 dt' \, \alpha(t')I^l_i(t-t'). \label{eq_ia}
\end{align}
$I^l_i(t)$ is derived from the sum of synaptic connections in which spikes are fired.
\begin{align}
I^l_i(t) &= \sum^N_{j=1} J^{l-1}_{ji} \sum^n_{k = 1} \delta(t - t^{l-1}_{j, k}),
\end{align}
where $t^l_{i,k}$ indicates the times that the $i$th neuron on layer $l$ fires.  $\beta$ is a conversion constant.

Membrane potential dynamics follow the spike-and-reset rule; when the membrane potential $v^l_i(t)$ reaches the threshold $V_{\mathrm{th}}$, a spike is fired, and after the absolute refractoriness $t_{\mathrm{ref}}$, the membrane potential is reset to the resetting potential $V_{\mathrm{reset}}$.

For the following analysis, we introduce the order parameter function $m^{l,\mu}(t)$, namely the overlap, defined by
\begin{align}
m^{l, \mu}(t) = \frac{2}{N} \sum^N_{i=1} \xi^{l, \mu}_i \sum^n_{k = 1} \delta(t - t^l_{i, k}).\label{eq_m}
\end{align}
Here, the overlap means how much the firing pattern matches the $\mu$th memory pattern on layer $l$.  If neurons with memory patterns $\xi^{l,\mu}_i=+1$ fire once, then $\int^{\infty}_{-\infty}dt \, m^{l,\mu}(t)=1$.  By using the overlap, $I^l_i(t)$ can be rewritten as
\begin{align}
I^{l}_i(t) = \frac{1}{2} \sum^p_{\mu=1} \xi^{l, \mu}_i m^{l-1, \mu}(t).\label{eq_i}
\end{align}
This means that the synaptic current to a neuron depends only on the overlap of the preceding layer and its memory patterns.

Throughout this paper, the parameter values are fixed as follows:  $N=1000$, $p=3$, $V_{\mathrm{rest}}=V_{\mathrm{reset}}=0$ mV, $V_{\mathrm{th}}=15$ mV, $t_{\mathrm{ref}}=1$ ms, $\tau=10$ ms, $I_0=0.075$ pA, $C=100$ pF, $D'=1$, $\alpha=2$ ms$^{-1}$, and $\beta=0.34$ pA.
\section{Fokker-Planck Method}\label{sec3}
In this section, we introduce the stochastic analysis of the membrane potential.  First, we define a vector whose elements are memory patterns of the $i$th neuron as $\bm{\xi^l_i} = ({\xi^{l,1}_i, \xi^{l,2}_i, \ldots ,\xi^{l,p}_i})$.  Each element takes on a value $+1$ or $-1$, and thus this vector has $2^p$ combinations.  We can define $2^p$ groups according to $\bm{\xi}^l_i$ values.  We call each group a sublattice and we discriminate each sublattice with the vector $\bm{\xi} = ({\xi^{1}, \xi^{2}, \ldots ,\xi^{p}})$.  Each element $\xi^{\mu}$ takes on $\pm1$ values.

Neurons belonging to the same sublattice receive the same synaptic current, because the synaptic current depends on only the overlaps and its memory pattern $\bm{\xi}^l_i$ (eq. (\ref{eq_i})).  The distribution of the membrane potential is known to evolve according to the Fokker-Planck equation,~\cite{cateau}
\begin{align}
\frac{\partial}{\partial t}P^l_{\bm{\xi}}(v, t) &= -\frac{\partial}{\partial v}J^l_{\bm{\xi}}(v,t) + \delta(v-V_{\mathrm{reset}})2^{-p}\nu^l_{\bm{\xi}}(t-t_{\mathrm{ref}}),\label{eq_fp}
\end{align}
where, 
\begin{align}
J^l_{\bm{\xi}}(v,t) &= -\left( \frac{v-V_{\mathrm{rest}}}{\tau}-\frac{I^{l, \alpha}_{\bm{\xi}}(t)+\mu}{C}+\frac{\partial}{\partial v}\frac{{D'}^2}{2} \right) P^l_{\bm{\xi}}(v,t).\label{eq_j_fp}
\end{align}
$P^l_{\bm{\xi}}(v,t)$ is the distribution of the membrane potentials of the neurons belonging to sublattice $\bm{\xi}$, and $\nu^l_{\bm{\xi}}(t)=2^p J^l_{\bm{\xi}}(V_{th},t)$ is the firing rate.  $P^l_{\bm{\xi}}(v,t)$ and $\nu^l_{\bm{\xi}}(t)$ satisfy the normalization condition,
\begin{align}
\int^{V_{th}}_{-\infty}dv \, P^l_{\bm{\xi}}(v,t) + 2^{-p}\int^{t_{\mathrm{ref}}}_{0} d \tau \, \nu^l_{\bm{\xi}}(t-\tau) =2^{-p}.
\end{align}

Here we define the overlap vector, $\bm{m}^l(t) = (m^{l,1}(t), m^{l,2}(t), \ldots, m^{l,p}(t) )$.  From eqs.~(\ref{eq_ia}) and (\ref{eq_i}), we can describe the synaptic current $I^{l, \alpha}_{\bm{\xi}}(t)$ by using $\bm{\xi}$ and $\bm{m}^l(t)$ as follows:
\begin{align}
I^{l, \alpha}_{\bm{\xi}}(t) &= \beta \int^{\infty}_0 dt' \, \alpha(t)I^l_{\bm{\xi}}(t-t'),\label{eq_ia_fp}\\
I^l_{\bm{\xi}}(t) &= \frac{1}{2} \bm{\xi} \cdot \bm{m}^{l-1}(t) .\label{eq_i_fp}
\end{align}
From eq.~(\ref{eq_m}), we can describe the overlap $m^{l,\mu}(t)$ by using firing rate $\nu^l_{\xi}(t)$ as 
\begin{align}
m^{l,\mu}(t) = 2^{1-p} \left ( \sum_{\xi^{\mu}=+1} \nu^l_{\bm{\xi}}(t) - \sum_{\xi^{\mu}=-1} \nu^l_{\bm{\xi}}(t) \right) .\label{eq_m_fp}
\end{align}
Here we can describe the network dynamics only by using macroscopic parameters $P^l_{\bm{\xi}}(v,t)$, $\nu^l_{\bm{\xi}}(t)$, $\bm{m}^l(t)$, and $I^l_{\bm{\xi}}$.

The description with the Fokker-Planck method is consistent with the LIF simulation in the limit of the number of neurons belonging to each sublattice going to infinity, $N/2^p \to \infty$.  Therefore, we restrict the total number of memory patterns to $p \sim O(1)$.
\section{Results}\label{sec4}
\subsection{Activation of a single memory pattern}\label{sec41}
\begin{figure}[tb]
\centering
\subfigure[$m=0.6$]{\includegraphics[width=7cm]{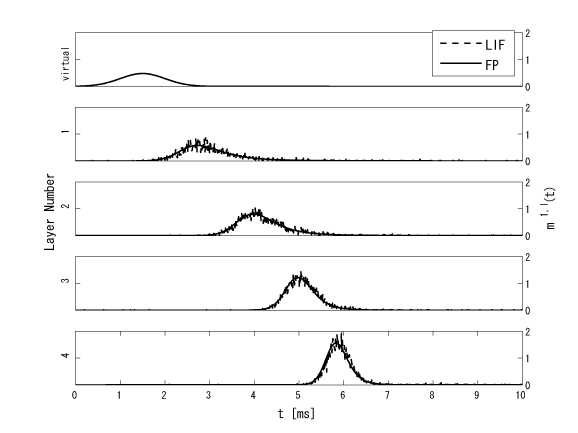}}
\subfigure[$m=0.4$]{\includegraphics[width=7cm]{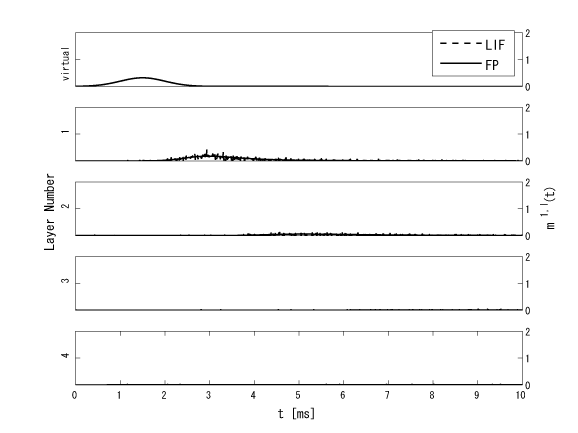}}
\caption{Overlaps $m^{l,1}(t)$ $(l=0\cdots4)$ when the total volume of the overlap on the virtual layer is set to $m=0.6$(a) or $m=0.4$(b), and the standard deviation is set to $\sigma=0.5$ ms.  The dashed line is obtained with the LIF simulation, and the solid line with the Fokker-Planck method.  We define $t_0$, when $m^{0,1}(t)$ takes a peak value, as $3\sigma = 1.5$ ms.}
\label{fig_single}
\end{figure}

Here we answer the question of whether each memory pattern synchronously propagates in this network as a pulse packet by using the LIF simulation and the Fokker-Planck method.  We activate the first layer of the network.  The initial condition is a stationary distribution for no external input, $I^l_i(t)=0$.  We use the overlap as an index of how firing patterns match the memory pattern. For the first layer activation, we consider the virtual layer and describe the overlap on the virtual layer of the first memory pattern as a Gaussian function with standard deviation $\sigma$ and total volume $m$. The volumes of the other memory patterns are set to 0;
\begin{align}
m^{0, \mu} (t)= \left\{
	\begin{array}{cl}
	\dfrac{m}{\sqrt{2 \pi} \sigma} \exp \left( \dfrac{(t-t_0)^2}{2 {\sigma}^2} \right) & \mu = 1,\\
	0 & \mu \neq 1.
	\end{array}
\right.\label{eq_input}
\end{align}
We calculate the input currents to neurons from eqs.~(\ref{eq_ia}) and (\ref{eq_i}), membrane potential dynamics from eq.~(\ref{eq_lif}), and the overlaps of the first memory pattern $m^{l,1}(t)$ from eq.~(\ref{eq_m}).  Figures~\ref{fig_single}(a) and \ref{fig_single}(b) plot the overlaps as dashed lines.  Each figure shows the overlaps of the five layers vertically.  Figure~\ref{fig_single}(a) is the case that the total volume of the overlap of the virtual layer is set to $m=0.6$, and Fig.~\ref{fig_single}(b) is the case with $m=0.4$.  In Fig.~\ref{fig_single}(a) the temporal profile of the overlap becomes sharper as the activity propagates through the layers.  Therefore, the memory pattern synchronously propagates as a pulse packet under this condition.  In contrast, in Fig.~\ref{fig_single}(b), the overlap dies out as the activity propagates.  That is, the memory pattern does not propagate under the latter condition.

Next, we calculate the input current to the neurons belonging to each sublattice from eqs.~(\ref{eq_ia_fp}) and (\ref{eq_i_fp}), membrane potential distributions and firing rates from the Fokker-Planck equation (eq.~(\ref{eq_fp})), and the overlaps of the first memory pattern $m^{l,1}(t)$ from eq.~(\ref{eq_m_fp}).  Figures~\ref{fig_single}(a) and \ref{fig_single}(b) plot the overlaps as solid lines.  Since the overlaps of the $\mu(\neq 1)$th memory pattern are 0, it is enough to divide the neurons into the sublattices only with the first memory pattern.  Therefore, we consider two distributions on each layer, $P^l_+$ and $P^l_-$;  $P^l_+$ is the membrane potential distribution of the $\bm{\xi}^l = (+1)$ sublattice, and $P^l_-$ is that of the $\bm{\xi}^l = (-1)$ sublattice.  Figures~\ref{fig_single}(a) and \ref{fig_single}(b) show the consistency between the results of the LIF simulation and those of the Fokker-Planck method.

\begin{figure}[tb]
\centering
\includegraphics[width=7cm]{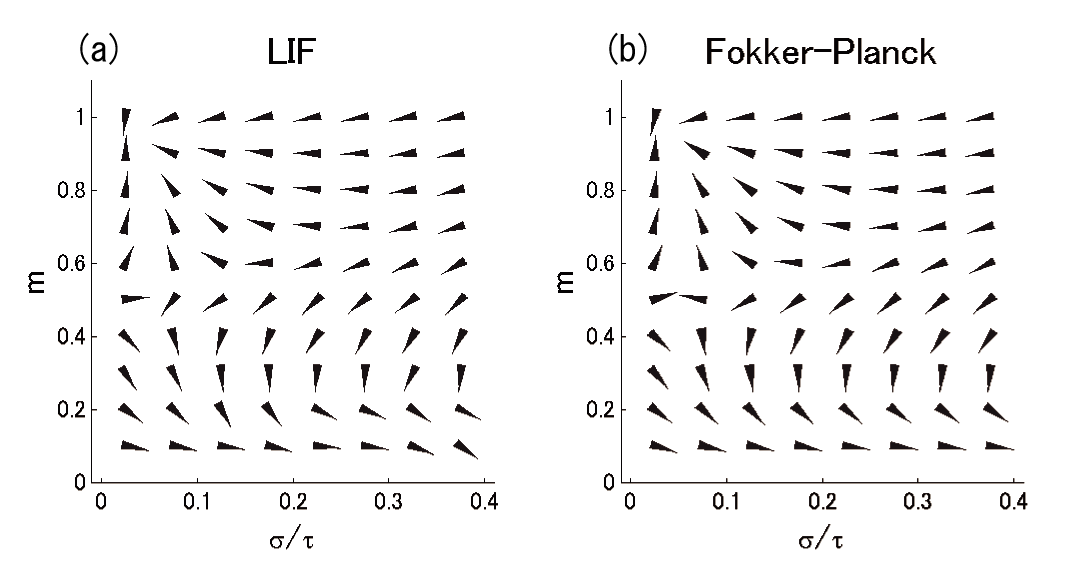}
\caption{Each arrow means the change from an input overlap to an output overlap.  The input overlaps are Gaussian functions, and the output overlaps are approximated as Gaussian functions. (a) LIF simulation result, (b) Fokker-Planck method result.}
\label{fig_vector}
\end{figure}

We choose the overlap of the virtual layer to be a Gaussian function with total volume $m$ and standard deviation $\sigma$.  Now, we evaluate the total volume $m'$ and the standard deviation $\sigma'$ of the overlap of the first layer by approximating the overlap with a Gaussian function by using the method of least squares.  Figure~\ref{fig_vector} shows the normalized vector from $(m, \sigma)$ to $(m', \sigma ')$ in.  Figure \ref{fig_vector}(a) shows the results of the LIF simulations and Fig.~\ref{fig_vector}(b) shows those of the Fokker-Planck method.  Figure \ref{fig_vector} shows that if the initial input is strong enough and synchronous, the overlap reaches the attractor on the left.  This attractor means that the spike pattern matches the memory pattern and the spikes are synchronized.  It indicates that this network works as a synfire chain and the memory pattern propagates as a pulse packet.  In contrast, when the initial input is too weak or too dispersed, the overlap dies out in the lower right region.  The results of the LIF simulation (Fig.~\ref{fig_vector}(a)) are consistent with those of the Fokker-Planck method (Fig.~\ref{fig_vector}(b)).

Hereafter we show only the results of the Fokker-Planck method, as all of the results are consistent with the LIF simulations.
\subsection{Activation of two memory patterns}\label{sec42}
\begin{figure}[tb]
\centering
\includegraphics[width=6cm]{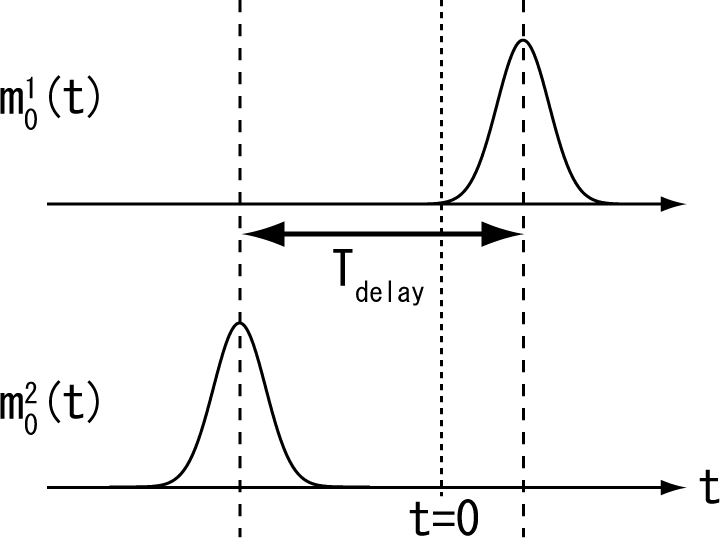}
\caption{Schematic diagram of the activation of two memory patterns.  We activate the first memory pattern $T_{\mathrm{delay}}$ after the second memory pattern.  We define the time of the onset of the first memory pattern activation as 0.}
\label{fig_delay}
\end{figure}

In the preceding subsection, we showed that in a layered associative network, a memory pattern could propagate as a spike packet.  Here we focus on the dynamics when several memory patterns are activated.  We analyze the simplest case, the activation of two memory patterns with an arbitrary interval $T_{\mathrm{delay}}$.  Figure~\ref{fig_delay} is a schematic diagram of pattern activation.  Here we consider two situations: simultaneous activation $T_{\mathrm{delay}}=0$ and successive activation $T_{\mathrm{delay}}>0$.  In both situations, we divide neurons into sublattices according to the signs of the patterns, because the overlaps of unfocused memory patterns are 0.  The sublattices $\bm{\xi}= (+1, +1), (+1,-1), (-1,+1),$ and $(-1,-1)$ are described as $(++), (+-), (-+)$, and $(--)$, and the membrane potential distributions are accordingly divided into the following groups: $P^l_{++}, P^l_{+-}, P^l_{-+},$ and $P^l_{--}$. We denote the firing rates of each sublattice as $\nu^l_{++}, \nu^l_{+-},\nu^l_{-+},$ and $\nu^l_{--}$.

\subsubsection{Simultaneous Activation of two memory patterns}\label{sec421}
Let us consider the case in which two memory patterns are activated simultaneously.  The overlaps of the virtual layer are described as follows:
\begin{align}
m^{0, \mu} (t)= \left\{
	\begin{array}{cl}
	\dfrac{m^\mu}{\sqrt{2 \pi} \sigma^\mu} \exp \left( \dfrac{(t-t_0^{\mu})^2}{2 {\sigma^\mu}^2} \right) & \mu = 1,2,\\
	0 & \mu \neq 1,2.
	\end{array}
\right.\label{eq_2input}
\end{align}
Regarding simultaneous activation, we assume that neurons on the virtual layer fire once or not at all.  The sum of total volumes of overlaps satisfies $m^1+m^2 \le 1$ because we consider two orthogonal memory patterns.  Here, we stipulate that $m^1+m^2=1, \sigma^1=\sigma^2=0.5$ ms, and $t^1_0 = t^2_0$.  We change only the ratio of $m^1$ to $m^2$.

First, we analyzed the simplest situation; $m^1 = 1$ and $m^2=0$.  As we have seen before, the overlap of the second memory pattern was always 0, and the first memory pattern propagated as a pulse packet as in Fig.~\ref{fig_single}(a).

\begin{figure}[tb]
\centering
\includegraphics[width=7cm]{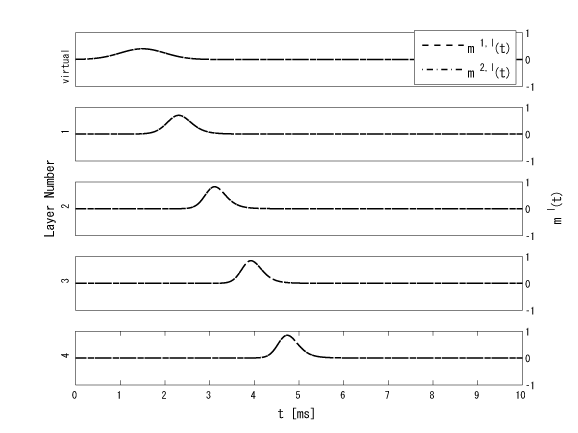}
\caption{Overlaps $m^{l,\mu}(t)$ when two memory patterns are activated with the same strength; $m_1 = m_2 = 0.5$.  We define $t^1_0$, when $m^{0,1}(t)$ takes a peak value, as $3\sigma^1 = 1.5$ ms.}
\label{fig_mix5}
\end{figure}

Second, we analyzed balanced activation; $m^1 = m^2 = 0.5$.  Figure \ref{fig_mix5} shows the temporal profiles of the overlaps.  Throughout the observation, both overlaps $m^{l,1}(t), m^{l,2}(t)$ have the same value and their total volumes were $\int^{\infty}_{-\infty} dt \, m^{\mu,l}(t) \sim 0.5$.  This situation seems to indicate that both memory patterns propagate with their intermediate levels.  We call this state a mixed state.

\begin{figure}[tb]
\centering
\includegraphics[width=7cm]{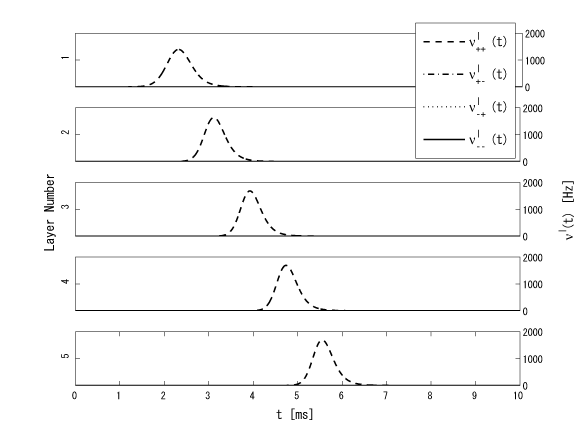}
\caption{Firing rate $\nu^l_{\bm{\xi}}(t)$ of each sublattice when two memory patterns are simultaneously activated with the same strength($m_1 = m_2 = 0.5$).}
\label{fig_mix5_fire}
\end{figure}

In order to elucidate which sublattices contain firing neurons, we focus on the firing rate of each sublattice.  Figure~\ref{fig_mix5_fire} shows the firing rates $\nu^l_{\bm{\xi}}$.  This figure shows that a mixed state is one in which spikes of $\bm{\xi}=(++)$ propagate.  Although the $\bm{\xi}=(++)$ sublattice is not a memory pattern, the activity of $\bm{\xi}=(++)$ propagates as a pulse packet.  Therefore, a layered associative network is a synfire chain in which not only memory patterns but also their mixed states propagate as pulse packets.

\begin{figure}[tb]
\centering
\includegraphics[width=7cm]{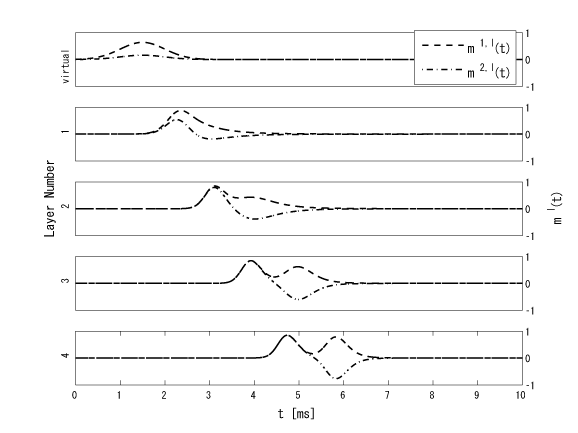}
\caption{Overlaps $m^{l,\mu}(t)$ when the first memory pattern is more strongly activated than second memory pattern ($m_1=0.8$, $m_2=0.2$).}
\label{fig_mix8}
\end{figure}

Next, we gradually increase $m^1$ from 0.5 to 1 while stipulating that $m^1 + m^2=1$.  As long as $m^1$ is approximately smaller than 0.7, $m^{1,l}(t) > m^{2,l}(t)$ early in several layers.  Despite this, the network finally converges to the mixed state, as in Figs.~\ref{fig_mix5} and \ref{fig_mix5_fire}.  In contrast, when $m^1$ is larger than a certain threshold, $m^{1,l}(t)$ has two peaks, and $m^{2,l}(t)$ has one positive peak and one negative peak.  Figure~\ref{fig_mix8} plots the overlaps for $m^1 = 0.8$ and $m^2=0.2$.  After convergence, $\int^{\infty}_{-\infty} dt \, m^{1,l}(t) \sim 1$ and $\int^{\infty}_{-\infty} dt \, m^{2,l}(t) \sim 0$.  In this regard, this situation is similar to when only the $\mu=1$ memory pattern is activated.  However, it is different in that the overlaps have two peaks.  We call this state a two-peak state.

\begin{figure}[tb]
\centering
\includegraphics[width=7cm]{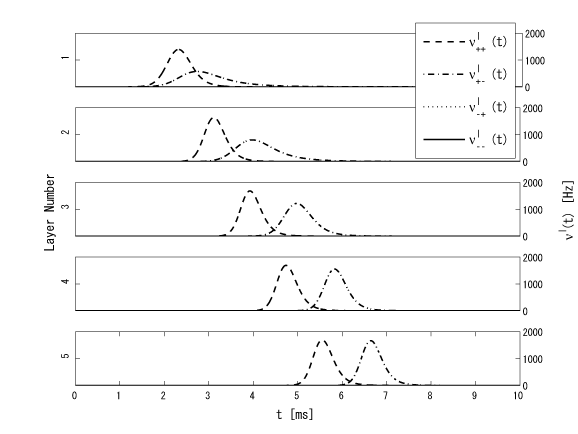}
\caption{Firing rate $\nu^l_{\bm{\xi}}(t)$ of each sublattice when the first memory pattern is more strongly activated than the second memory pattern; $m_1=0.8$, $m_2=0.2$.}
\label{fig_mix8_fire}
\end{figure}

Similar to Fig.~\ref{fig_mix5_fire}, Fig.~\ref{fig_mix8_fire} shows the firing rate of each sublattice for the case of the two-peak state.  Figure \ref{fig_mix8_fire} indicates that neurons in $\bm{\xi}=(++)$ and $(+-)$ sublattices fire at different timings.  This difference is due to the difference in synaptic current strengths.  In the first layer, $I_{++}(t) = \frac{1}{2}(m^{1,l}(t) + m^{2,l}(t)) > I_{+-}(t) = \frac{1}{2}(m^{1,l}(t) - m^{2,l}(t))$ because $m^{2,l}(t) >0$.  Therefore, neurons in the $(++)$ sublattice receiving the larger current fire earlier than those in the $(+-)$ sublattice.

The timing difference does not grow after convergence in the two-peak state.  To elucidate the reason, we rewrite $I^{l+1}_{\bm{\xi}}$ as follows:
\begin{align}
I^{l+1}_{++} &= \frac{1}{2}\left(m^{1,l}(t) + m^{2,l}(t) \right)\nonumber \\
&= \frac{1}{4} \Bigl( \nu^l_{++}(t) + \nu^l_{+-}(t) - \nu^l_{-+}(t) - \nu^l_{--}(t) \nonumber \\
&\qquad {} + \nu^l_{++}(t) + \nu^l_{-+}(t) - \nu^l_{+-}(t) - \nu^l_{--}(t) \nonumber \Bigr)\\
&= \frac{1}{2} \left( \nu^l_{++}(t) -\nu^l_{--}(t) \right), \label{indep_s}
\end{align}
Similarly, 
\begin{align}
I^{l+1}_{+-} &= \frac{1}{2} \left( \nu^l_{+-}(t) -\nu^l_{-+}(t) \right),\\
I^{l+1}_{-+} &= \frac{1}{2} \left( \nu^l_{-+}(t) -\nu^l_{+-}(t) \right) = -I^{l+1}_{+-},\\
I^{l+1}_{--} &= \frac{1}{2} \left( \nu^l_{--}(t) -\nu^l_{++}(t) \right) = -I^{l+1}_{++}.\label{indep_e}
\end{align}
Therefore, the activities of the $\bm{\xi}=(++),(--)$ sublattices independently propagate from $(+-), (-+)$ sublattices.  Note that this independence cannot be achieved for $p \ge 3$ cases.  In this $p=2$ situation, the activity of the $(++)$ sublattice independently propagates from that of $(+-)$.  After each activity reaches a synchronous state, which is the attractor in Fig.~\ref{fig_vector}, both activities propagate at the same speed.

This analysis of firing rate dynamics indicates that the success or failure of propagation of the $(+-)$ sublattice activity can be represented as a two-peak state or a mixed state, respectively.

Finally, we gradually increased $m^1$ to 1.  As $m^1$ became larger, the timing difference between the two peaks became smaller and finally vanished when $m^1 = 1$.  Thus, we can consider the single pattern propagation as a special case of a two-peak state.

In summary, simultaneous activation under the restriction of $m^1+m^2=1$ gives rise to two states, a mixed state and a two-peak state.  If $|m^1-m^2|$ is small enough, the network converges to a mixed state; otherwise it converges to a two-peak state. 
\subsubsection{Successive Activation of two memory patterns}\label{sec422}
C\^ateau and Fukai reported that when they activated two pulse packets successively with a short interval, the following pulse packet did not propagate because of hyperpolarization caused by the preceding pulse packet's propagation.\cite{cateau}

Let us consider the case in which two memory patterns are successively activated with a short interval, $T_{\mathrm{delay}}>0$.  Here, because of the time interval, we do not restrict the number of spikes, which is in contrast to the simultaneous activation; thus $m^1+m^2$ can be more than 1.  For simplicity, we assume that the overlaps of the virtual layer of both memory patterns take on the same values except for the timing, and that the parameters of the Gaussian functions describing the virtual layer overlaps are $m^1=m^2 = 0.7$, and $\sigma^1=\sigma^2 = 0.5$ ms in eq. (\ref{eq_2input}).  We activate the following $\mu=1$ memory pattern following the preceding $\mu=2$ memory pattern (Fig.~\ref{fig_delay}).

When $T_{\mathrm{delay}}$ is much larger than the membrane time constant $\tau=10$ ms, for example $T_{\mathrm{delay}} = 50$ ms, the $\bm{\xi}=(++)$ and $(-+)$ sublattices are simultaneously activated by the preceding $\mu=2$ activation, and the $\bm{\xi}=(++)$ and $(+-)$ sublattices are almost simultaneously activated by the following $\mu=1$ activation (data not shown).  In this situation, the following memory pattern seemed to propagate normally without being affected by the preceding memory pattern's propagation.

\begin{figure}[tb]
\centering
\includegraphics[width=7cm]{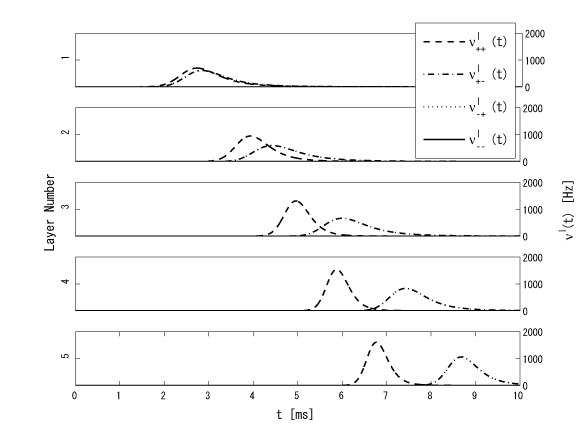}
\caption{Firing rate of each sublattice when we activate the first memory pattern $T_{\mathrm{delay}}=20$ ms after the second memory pattern.  $t^1_0$, when $m^{0,1}(t)$ has a peak value, is defined as $3\sigma^1 = 1.5$ ms.}
\label{fig_d20_fire}
\end{figure}

However, when the interval $T_{\mathrm{delay}}$ is decreased, the spikes of neurons belonging to $\bm{\xi}=(+-)$ sublattice are gradually delayed compared with those of $(++)$ neurons during the following pattern propagation.  Figure~\ref{fig_d20_fire} shows the firing rate of each sublattice when the interval $T_{\mathrm{delay}} = 20$ ms.  This state is two-peaked.

\begin{figure}[tb]
\centering
\includegraphics[width=6cm]{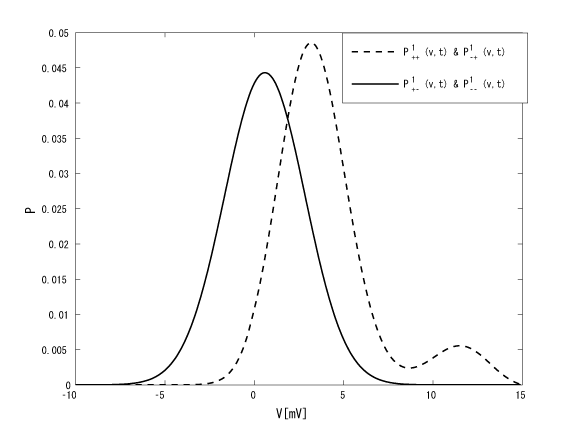}
\caption{Membrane potential distribution of the first layer of each sublattice after propagation of the second memory pattern, 6.5 ms after $t^2_0$.  The membrane potential distribution of the $\bm{\xi}=(++),(-+)$ sublattice is plotted as the dashed line and that of $(+-),(--)$ as the solid line.}
\label{fig_vdist}
\end{figure}

To investigate the reason for this delay, we plot the membrane potential distribution of the first layer of each sublattice $P^l_{\bm{\xi}}$ after the propagation of the preceding memory pattern in Fig.~\ref{fig_vdist}.  During propagation of the preceding memory pattern, neurons in the $\bm{\xi}=(++)$ and $(-+)$ sublattices fire and their membrane potential are reset to $V_{\mathrm{reset}}$ (Fig.~\ref{fig_vdist} dashed line).  In contrast, neurons in the $(+-)$ and $(--)$ sublattices hyperpolarize because of inhibitory current (Fig.~\ref{fig_vdist}, solid line).  Therefore when the input from the following memory pattern arrives, neurons in the $(+-)$ sublattice take more time than the ones in the $(++)$ sublattice to reach the threshold $V_{th}$.  This is the origin of the delay.

In the deeper layers, a once dispersed $(+-)$ pulse packet develops into a synchronized pulse packet, and the interval between the preceding pulse packet and $(+-)$ pulse packet approaches a certain value, which is long enough for the membrane potential distribution of the $(+-)$ sublattice to relax to the stationary state.  These dynamics are independent of the $(++)$ sublattice dynamics.  The $(++)$ pulse packet also develops into the synchronized state, and the interval between the preceding pulse packet and the $(++)$ pulse packet also converges to a certain interval.  Therefore, the delay between $(++)$ and $(+-)$ also converges to a certain delay period.

If the time interval is further decreased, the activity of the $(+-)$ sublattice does not propagate after the following memory pattern activation.  Figure \ref{fig_d15_fire} shows the firing rate of each sublattice when the time interval is $T_{\mathrm{delay}}=15$ms.  This figure shows that only the activity of the $(++)$ sublattice propagates whereas that of $(+-)$ dies out.  This is the mixed state described in Fig.~\ref{fig_mix5_fire}.  The activity of $(+-)$ does not propagate because the smaller $T_{delay}$ is, the more the inhibitory effect remains and it is more difficult for neurons belonging to $(+-)$ to fire when the following memory pattern is activated.

\begin{figure}[tb]
\centering
\includegraphics[width=7cm]{delay_fire15.png}
\caption{Firing rate of each sublattice when the first memory pattern $T_{\mathrm{delay}}=15$ ms is activated after the second memory pattern.}
\label{fig_d15_fire}
\end{figure}

When the time interval is decreased even further, for example $T_{\mathrm{delay}} = 8$ms, even the activity of the $(++)$ sublattice does not propagate (data not shown).  This is because of hyperpolarization after spikes in the preceding activation.  Therefore it is difficult even for neurons belonging to the $(++)$ sublattice to fire when the first memory pattern is activated.  This phenomenon is similar to the one reported by C\^ateau and Fukai \cite{cateau}.
\section{Summary and Discussion}
\label{sec5}
In this paper, we considered a model of a layered associative network constructed by LIF neurons and analyzed it with the Fokker-Planck method.  We showed that mixed states, not only memory patterns, propagated as pulse packets through the network.  When we activated a memory pattern much more strongly than another memory pattern, we observed a characteristic phenomenon in which the overlaps have two peaks.  On successive activations, the network converges to a two-peak state, mixed state, or non-spike state depending on the interval duration.

The difference between the conventional Ising neuron network and our network is the stability around memory patterns.  The conventional network converges to a memory state or a mixed state and not to a two-peaked state.  The two-peaked state is a characteristic state for neurons which integrate inputs to fire because the neurons driven by different input strengths fire at different timings.  In contrast, binary neurons fire simultaneously for uneven current magnitudes as long as the current reaches the threshold.  For example, in Fig.~\ref{fig_mix8_fire}, binary neurons belonging to $\bm{\xi}=(++)$ and $(+-)$ fire simultaneously, and this means the network is in a $\mu=1$ memory state.  In the LIF network, the timing difference conveys some information about input balance and timing.

\section*{Acknowledgments}
This work was partially supported by a Grant-in-Aid for Scientific Research on Priority Areas No. 14084212, and for Scientific Research (C) No. 16500093 from the Ministry of Education, Culture, Sports, Science and Technology of Japan.

\end{document}